\documentclass[prl,twocolumn,superscriptaddress,letterpaper,floatfix]{revtex4}
\usepackage{graphicx}

\begin{document}
\title{Nitrogen based magnetic semiconductors}

\author{I.S.~Elfimov}
\affiliation{Department of Physics and
Astronomy, University of British Columbia, 6224 Agricultural
Road, Vancouver, British Columbia, Canada, V6T 1Z1}

\author{A.~Rusydi} 
\affiliation{Institut f\"{u}r Angewandte Physik,
Univerist\"{a}t Hamburg, Jungiusstra$\ss$e 11, D-20355 
Hamburg, Germany}

\author{S.I.~Csiszar} 
\affiliation{Chemical Physics
Laboratory, Materials Science Centre, University of
Groningen, Nijenborgh 4, 9747 AG Groningen, The Netherlands}

\author{Z.~Hu} \affiliation{II. Physikalisches Institut, 
Univerist\"{a}t zu K\"{o}ln, 
Z\"{u}lpicher Str. 77, D-50937 K\"{o}ln, Germany}

\author{H.~H.~Hsieh} \affiliation{National Synchrotron 
Radiation Research Center, 
101 Hsin-Ann Road, Hsinchu 30077, Taiwan}

\author{H.-J.~Lin} \affiliation{National Synchrotron 
Radiation Research Center, 
101 Hsin-Ann Road, Hsinchu 30077, Taiwan}

\author{C.T.~Chen} \affiliation{National Synchrotron 
Radiation Research Center, 
101 Hsin-Ann Road, Hsinchu 30077, Taiwan}

\author{R.~Liang} \affiliation{Department of Physics
and Astronomy, University of British Columbia, 6224
Agricultural Road, Vancouver, British Columbia, Canada, V6T
1Z1}

\author{G.A.~Sawatzky} \affiliation{Department of Physics
and Astronomy, University of British Columbia, 6224
Agricultural Road, Vancouver, British Columbia, Canada, V6T
1Z1}

\begin{abstract}
We describe a possible pathway to new magnetic materials with 
no conventional magnetic elements present. 
The substitution of Nitrogen for Oxygen in simple non 
magnetic oxides leads to holes in N 2$p$ states which form 
local magnetic moments. Because of the very large Hund's rule
coupling of Nitrogen and O 2$p$ electrons and the rather extended 
spatial extend of the wave functions these materials are 
predicted to be ferromagnetic metals or small band gap 
insulators. Experimental studies support the theoretical 
calculations with regard to the basic electronic structure 
and the formation of local magnetic moments. It remains to be 
seen if these materials are magnetically ordered and if so 
below what temperature. 
\end{abstract}


\maketitle

Since the discovery of ferromagnetism in Mn-doped
GaAs\cite{ohno98} the study of 
magnetic impurities in  non magnetic traditional semiconductors
become a major
trend in the field called diluted magnetic semiconductors
(DMS). Recently it was proposed that oxides
based DMS form another class of materials with promise
for ferromagnetism at temperatures even higher than are
currently achieved with III-IV semiconductors\cite{dietl00}. 
Following this prediction, several ZnO, and TiO$_2$, 
based systems have been reported to show ferromagnetism
at room temperature if substituted with 3$d$
elements\cite{matsumoto01, chambers01}. 

Nevertheless, in spite of accelerated interest
in these materials, the exact mechanism responsible
for the observed magnetic properties continues to be
strongly debated in the literature. 
Several studies reported
strong evidence of phase separation and formation
of ferromagnetic clusters suggesting a non intrinsic behavior
not suitable for technological applications. In
addition, some of ZnO:Mn and Co-doped systems were
reported to show only paramagnetic rather than ferromagnetic
properties\cite{fukumura01,yoon03,kolesnik04,park04}. 
It is believed that additional hole doping is needed to 
stabilize the ferromagnetic ground state in such 
DMS\cite{sato00,sato01}. Note that simple substitution
of Mn$^{2+}$ for divalent Zn in ZnO, for example, does
not generate any extra carriers in itself. This is in
contrast to Mn-doped GaAs where Mn introduces one
hole per substituted cation. The interaction between
these holes and electrons in 3$d$ impurity states is one
of the critical elements in the physics of diluted magnetic
semiconductors\cite{dietl00,okabayashi,krstajic}. 

\begin{figure} 
\centering
\includegraphics[clip=true,width=0.47\textwidth]{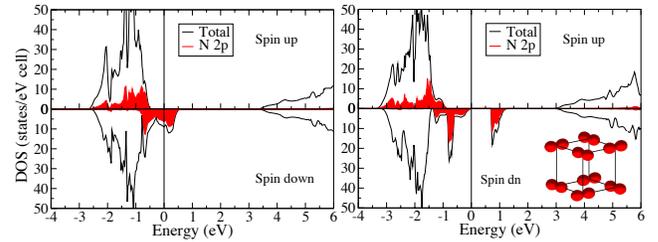}
\caption{Total and Nitrogen 2$p$ partial densities of
states calculated in LSDA (left panel) and LSDA+U (right
panel) for 25\% N substitution. LSDA+U density of states is
calculated with $U^{N2p}$=$U^{O2p}$=4.6eV, and
$J^{N2p}_{H}$=$J^{O2p}_{H}$=1.2eV. 
The zero of energy is at Fermi energy.} 
\label{fig:25pDOS} 
\end{figure}

Another important factor is the strong
Hund's rule coupling $J_H$ that favors a high-spin
configuration on the impurity site. Note that the large
splitting between occupied spin up and unoccupied spin
down states in Mn$^{2+}$ impurities is only partially due to
the Coulomb repulsion $U$ which is actually strongly reduced
in a solid. $J_H$, on the other hand, is hardly reduced
from its  atomic value of about 0.9eV per spin pair on average 
for the series of 3$d$ TM's\cite{marel88,solovyev96}. 
In this context, it is worthwhile to note that strong 
effective Hund's rule coupling between the carriers bound to 
vacancies was recently proposed as another way to 
introduce magnetism in none magnetic semiconductors\cite{CaO}.

Using cation vacancy in CaO as an example, it was
shown that two holes in otherwise full Oxygen 2$p$ orbitals
on the nearest neighbor sites of the vacancy sites form
a stable magnetic (spin triplet) ground state in a simple,
originally non magnetic band insulator. Note that even
at very small concentrations such charge compensating
clusters have significant overlap due to the extended nature
of these objects. 
The origin of such strong coupling
is, however, in the kinetic energy and symmetry of the
molecular orbitals rather than in local, on-site interactions.

Aside from this it is illuminating to realize that
the lowest energy configuration for two holes in 2$p$ orbitals
of atomic O or N is a spin triplet ($^3$P)
configuration and the energy required to reach the first
exited, spin singlet $^1$D, is 0.13Ry in the case of 
atomic O\cite{slater}. This corresponds to 1.47eV for $J_H$.
Note that the Hartree-Fock value of $J_H$ for 3$d$ 
electrons in Cu, for example, is 1.45eV\cite{marel88}. 
This demonstrates that there is no reason why a system 
with holes in anion $p$ bands could not be magnetic 
provided the band width is not too large.  

\begin{figure} 
\centering
\includegraphics[clip=true,width=0.47\textwidth]{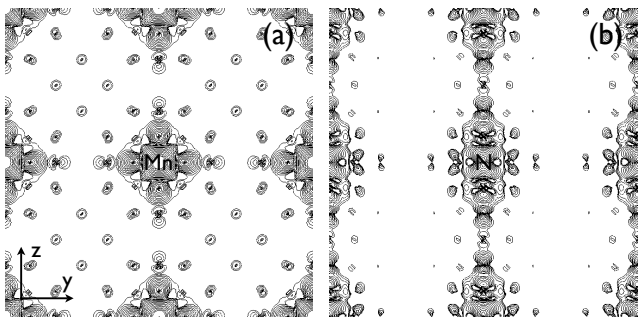}
\caption{LSDA+U spin density calculated for ~0.9\% Mn (a) and
Nitrogen (b) substitution in SrO with $U^{N2p}$=$U^{O2p}$=4.6eV,
$J^{N2p}_{H}$=$J^{O2p}_{H}$=1.2eV, $U^{Mn3d}$=2.34eV and 
$J^{Mn3d}_{H}$=0.89eV. Shown is a cut in (100) plane
of 6x6 super cell. Note that charge-compensating hole resides in 
the Nitrogen $p_z$ orbital. Lattice sites are not shown for 
clarity.} 
\label{fig:spin_den}
\end{figure}

Auger spectroscopy studies of
simple oxides show that $F^0$ integral for two holes 
in an O 2$p$ orbital is 
4-7 eV\cite{Tjeng88, Altieri_thesis}. 
Note that with Auger spectroscopy of otherwise
boring oxides one can determine
the energies of the  final states with
two holes on an Oxygen ion. In fact it
turns out that because of Auger matrix
elements only the two singlet states
can be reached being the singlet $^1S$ and
the $^1D$ state whose energy splitting 
determines $F^2$\cite{Sawatzky_auger_ln, Fuggle}.
In the case of Cu$_2$O the $F^2$ integral is 
found to be 6eV that corresponds to 
1.2eV for $J_H$\cite{Tjeng89, Hao_thesis}.   

In spite of such striking similarities, there are several 
important difference between TM 3$d$ orbitals
and Oxygen 2$p$ in oxides. First the the Oxygen 2$p$ 
bands are usually full in ionic oxides leaving no room 
for unpaired spins. Even in the hole doped cuprates the 
density of O holes is small so that the effect of this 
Hund's rule coupling will be rather small. 
An interesting exception is CrO$_2$ in which Oxygen 
based bands cross the Fermi energy and indeed there is 
unpaired spin density in O\cite{CrO2}. 
Secondly the much larger spatial extend of the O 2$p$ 
orbitals which could facilitate longer range exchange 
interactions.
Note that $d$-$d$ hopping integrals ($t_{dd}$) scale as 
the inverse of the fifth power of inter-atomic distance 
whereas $t_{pp}\sim1/r^{2}$\cite{harrison}. 
The consequence of this is two fold. 
First the larger spatial extend will cause large band 
widths which will compete with the tendency towards 
ferromagnetism. This would point in a direction of larger 
lattice constants such as is the case in SrO when comparing 
to the 3$d$ transition metal oxides. Secondly the larger 
hopping integrals will cause more delocalization over 
neighboring atoms resulting in longer range exchange 
interactions. The first is unfavorable but the second 
favorable. To get a win-win situation in which both 
effects favor ferromagnetism we suggested the use of N 
substitution for O. N$^{2-}$ has one less electron and 
because of the lower binding energy the N 2$p$ states 
will be closer to the chemical potential than the O 2$p$ 
states forming a relatively narrow impurity band with 
one hole per N situated just above the chemical potential. 
We note that such a substitution has recently been discussed
in the context of ferromagnetism in Mn-doped ZnO, for
example, where it is used to increase the ferromagnetic
coupling between Mn impurities\cite{sato00,sato01}. 
In the present
paper we present theoretical and experimental evidence
that substitution of Nitrogen for Oxygen in simple band
insulators such as SrO is, in itself, sufficient to produce
DMS. This reveals a whole class of new materials
whose magnetic properties are determined solely by
the interaction between the carriers in $p$ rather than $d$ or
$f$ atomic shells.

\begin{figure} 
\centering
\includegraphics[clip=true,width=0.33\textwidth]{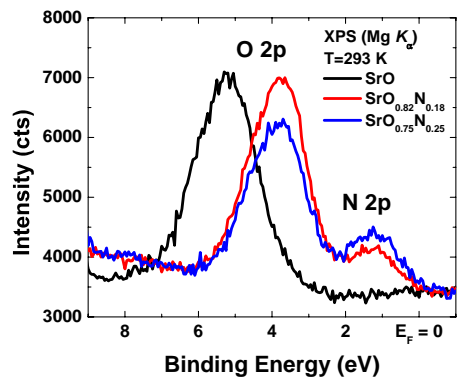} 
\caption{Room temperature XPS valence band spectra of
SrO$_{1-x}$N$_x$ films. The zero of binding energy is at
Fermi level of Copper.} 
\label{fig:XPS} 
\end{figure}

As many other divalent oxides, SrO has the rock salt
crystal structure with a lattice constant of 5.16\AA. 
Pure SrO is a nonmagnetic band insulator with a forbidden gap
of 5.3eV. The LDA-calculated gap is 3.9eV. In contrast
to TM in III-V semiconductors, Nitrogen substitution in
oxides can be substantially higher provided that one uses
molecular beam epitaxy (MBE) methods to introduce N. 
For example, Voogt {\it et al} reported that NO$_2$-assisted
MBE growth of FeO resulted in thin films with a rock salt
crystal structure where up to 18\% of anion sites were occupied
by Nitrogen\cite{voogt01}. It is, therefore, interesting to
investigate the effect of high Nitrogen concentration on
the properties of SrO$_{1-x}$N$_{x}$. 

Fig.\ref{fig:25pDOS} shows the total and
N 2$p$ partial density of states calculated for 25\% Nitrogen
substitution with TB-LMTO computer code\cite{tblmto} within 
the local spin density approximation (LSDA). 
As in the case of low N concentration, LSDA
predicts a ferromagnetic metallic ground state. However,
it is well known that LSDA usually fails to describe
strongly correlated systems adequately. In order
to account for correlation effects we performed LSDA+U
calculation\cite{ldau} with 
$U^{N2p}$=$U^{O2p}$=4.6eV, and
$J^{N2p}_{H}$=$J^{O2p}_{H}$=1.2eV.
It is found that SrO$_{0.75}$N$_{0.25}$ is, in fact a 
semiconductor with an energy gap of 0.7eV in ferromagnetic
spin arrangement (Fig.\ref{fig:25pDOS}) and 1.1eV 
in antiferromagnetic one. In addition, the ferromagnetic 
solution becomes metallic if $U$ is less than 3.3 eV 
whereas decreasing $J_H$ from 1.2 to 0.5eV 
in fact increases the gap by 0.3eV.
Note, however, that orbital degeneracy is lifted via the
mechanism of orbital ordering which is actually frustrated
if Nitrogen is distributed over the lattice in the
periodic fashion such as in a simple cubic super-structure.
One possible hole-density configuration, which corresponds
to the calculated ferromagnetic density of states,
is shown in the inset of Fig.\ref{fig:25pDOS}. 
We also note that, similar
to manganites, spin and orbital degrees of freedom in
these materials are strongly correlated which influences
drastically the low energy properties. Although, ferromagnetic
and A-type antiferromagnetic solutions are
degenerate within our computational approach (the total energy
difference between former and latter is 1meV), 
the G-type is 0.19eV higher in energy due to transition
from antiferro to ferro orbital ordering. This, of course
is very sensitive to lattice spacing, crystal structure distortion
and morphology of the actual material. It is,
therefore, really important to address these issues from
an experimental point of view.

To demonstrate the difference in the spatial extend of the 
induced spin density between transition
metal and ligand substitutions we compare first principle
density functional calculation of 0.9\% Mn and N substituted
in SrO. Within the bounds of LSDA+U approach, both systems 
are magnetic and insulating with the moment of 5 and 
1$\mu_B$ per supercell, respectively.  
We note that our LSDA calculations based on the supercell
approximation agree with  
coherent potential 
approximation results \cite{kenmochi}. The LSDA+U calculations of Mn doped
SrO were performed with $U^{Mn3d}$=2.34eV and $J^{Mn3d}_H$=0.89eV
\cite{marel88}. The values of $U$ and $J$ for O 2$p$ were
4.6 and 1.2eV, respectively. As one can see from 
Fig.\ref{fig:spin_den}, the spin density due to Mn is highly
concentrated around the first nearest Oxygen neighbors
whereas for Nitrogen it is spread out over several neighboring
shells and it is rather directional. This, of course,
directly follows from the spatial and angular dependences
of the hopping parameters for 2$p$ in comparison to 3$d$ orbitals.

\begin{figure} 
\centering
\includegraphics[clip=true,width=0.4\textwidth]{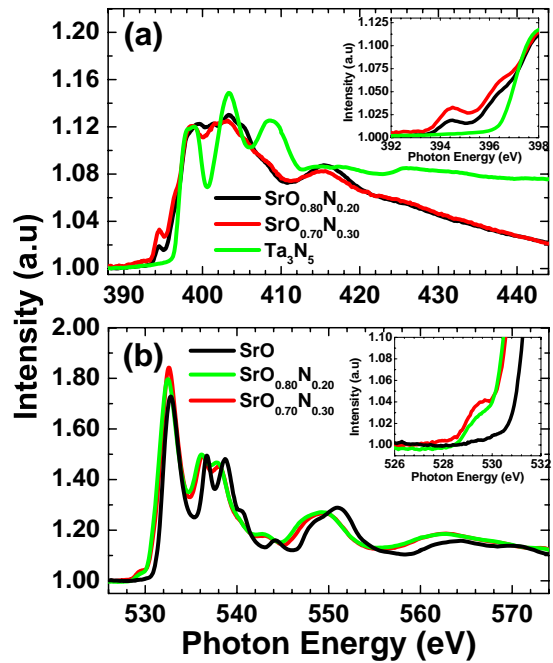} 
\caption{Room temperature x-ray absorption spectra 
with E$\bot c$ of SrO$_{1-x}$N$_x$ films,
taken in the total electron yield mode at the 
Nitrogen (a) and Oxygen (b) K-edges. 
Note pre-edge structure at 394.5eV in (a) and
529.25eV in (b).} 
\label{fig:XAS} 
\end{figure}

We tried to prepare SrO$_{1-x}$N$_x$ by solid state chemistry
methods using powders of SrN and SrO but found
in the end only phase separated materials or Sr Nitroxides.
However as mentioned above it is documented that
Fe Oxides contain substitutional Nitrogen if grown by
MBE methods using NO as the oxidizing gas. This suggests
that the lower temperatures used in MBE growth
together with the fact that only surface and not bulk
diffusion is needed for layer by layer growth results in
metastable materials with substitutional N. We note that
this may be of general importance in other new systems
such as phosphorous substituted sulfides etc. 

Films
were grown epitaxial on MgO (100) single crystal under Ultra
High Vacuum (UHV) conditions\cite{csiszar}. In order to produce
films of high quality we use Molecular Beam Epitaxy
(MBE). Reflection High Energy Diffraction (RHEED) was used to 
monitor the growth process.
The periodic oscillations observed in the specular spot
intensity of the high-energy electron beam reflected from
the film surface are evidence for a layer-by-layer type
of growth. Films with various concentrations of N were
grown using NO gas as an oxidizing agent. The structure
of the films was proven to be the same as the substrate
(FCC), although (due to a large lattice mismatch) after
several layers, having a different in-plane lattice constant
then the substrate. This in plane lattice constant turned
out to be also dependent on N concentration, namely
increasing with increasing N content. Good Low Energy
Electron Diffraction (LEED) image also confirms
the high quality of the films. A more detailed description
of the growth will be given elsewhere.

The composition of the films was checked by in
situ measurements of X-Ray Photoemission Spectroscopy
(XPS). In Fig.\ref{fig:XPS} we show XPS spectra taken at room
temperature on samples with 0\%, 18\% and 25\% Nitrogen.
In agreement with the results of band structure
calculations the N 2$p$ peak is found to be about 2eV
lower in binding energy relative to the position of O
2$p$ peak. The relative change in intensity of these two
peaks upon doping strongly indicates that the growth
process is indeed a process of substitution. This is also
supported by RHEED and LEED data. Note that the
shift of the O 2$p$ peak towards higher binding energy in
the pure SrO originates from charging problems common
for insulators\cite{hufner95}. 
A much smaller shift, if any, found
in the samples containing Nitrogen is an indication of
semiconducting or even metallic behavior. 

X-ray Absorption
Spectroscopy (XAS) measurements performed
on NSRRC Dragon Beamline in Taiwan confirm our XPS
results. This experiment was carried out with the incident
beam polarization parallel to the sample surface
(001 surface of MgO and SrO$_{1-x}$N$_x$). 
In Fig.\ref{fig:XAS} we show
XAS spectra taken in the total electron yield mode 
at the Nitrogen and Oxygen K-edges. 
The strong pre-peak at the N K-edge strongly 
supports the existence of N 2$p$ holes. 
The presence of a weaker O pre-edge structure 
supports the suggestion of delocalization of 
the N 2$p$ hole states onto neighboring O.
Note that neither Ta$_3$N$_5$ nor SrO has such
a feature in their K-edge spectra.

\begin{figure} 
\centering
\includegraphics[clip=true,width=0.4\textwidth]{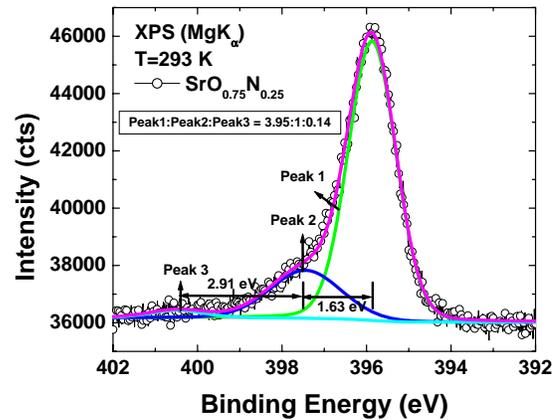} 
\caption{Room temperature N 1$s$ core-level XPS spectra
of $SrO_{1-x}N_x$ film with x~0.25. The three peaks
structure is due to a local spin 1/2 moment.}
\label{fig:coreXPS} 
\end{figure}

Nitrogen 1$s$ core-level
XPS was used to examine the degree of charge and spin
localization. If charge-compensating holes reside primarily
in the Nitrogen 2$p$ orbitals, the spectra of SrO$_{1-x}$N$_x$
films should be similar to one found in gas phase NO with
a double peak structure due to the exchange coupling of
the unpaired 2$p$ spin with the core-hole. If all other interactions
are frozen out then the intensity ratio between
a triplet (lower binding energy) and singlet (higher binding
energy) peaks should be 3:1. Of course, in the real
material this is often not the case and it is rather common
that experimental ratio is large\cite{st_ratio}. We find that
in the SrO$_{0.75}$N$_{0.25}$ sample the ratio is 4:1 
(Fig.\ref{fig:coreXPS}). The
observed energy separation between singlet and triplet
peaks is 1.6eV which is consistent with 1.5eV found in
NO which has a spin of 1/2 localized mainly on N. 

In conclusion, our experimental and theoretical results
show that substitution of Nitrogen for Oxygen in simple
nonmagnetic and insulating oxide results in a dramatic
change of the electronic and magnetic properties.
This opens a path to the new class of magnetic semiconductors
whose properties are solely determined by the
interactions between the holes in ligand p-orbitals. 

We acknowledge L.H.~Tjeng for for the technical assistance, 
use of the Cologne MBE and XAS system at NSRRC and 
stimulating discussions. 
The authors are grateful to T.~Hibma for use of the 
MBE and XPS systems as well as for the hospitality 
at the University of Groningen.
We also thank C.-C.~Kao and S.L.~Hulbert for valuable discussions. 
This work was supported by the Netherlands 
Organization for Fundamental Research on Matter (FOM), U.S. Department
of Energy contract DE-AC02- 98CH10886, as well
as the Canadian funding agencies NSERC, CIAR, and
CFI. Theoretical investigation has been enabled 
in part by the use of WestGrid computing resources.


\begin{references}

\bibitem{ohno98} H.~Ohno, Science {\bf 281} 951 (1998)

\bibitem{dietl00}  T.~Dietl, H.~Ohno, F.~Matsukura,
J.~Cibert, and D.~Ferrand, Science {\bf 287}, 1019 (2000)

\bibitem{matsumoto01} Y.~Matsumoto {\it et al.}, Science
{\bf 291}, 854 (2001) 

\bibitem{chambers01} S.~A.~Chambers {\it et al.} , Appl. Phys. Lett. 
{\bf 79}, 3467 (2001)

\bibitem{fukumura01} T.~Fukumura, {\it et al}, Appl. Phys. Lett. 
{\bf 78}, 958 (2001)
 
\bibitem{yoon03} S.W.~Yoon, {\it et al}, J. Appl. Phys. 
{\bf 93}, 7879 (2003)
 
\bibitem{kolesnik04} S. Kolesnik and B. Dabrowski,  J. Appl. Phys. 
{\bf 96}, 5379 (2004)

\bibitem{park04} Jung~H.~Park, {\it et al}, Appl. Phys. Lett. 
{\bf 84}, 1338 (2004)

\bibitem{sato00} K.~Sato and H.~Katayama-Yoshida, Jpn. J.
Appl. Phys. {\bf 39}, L555 (2000)

\bibitem{sato01} K.~Sato and H.~Katayama-Yoshida, Physica B
{\bf 308}, 904 (2001)

\bibitem{okabayashi} J.~Okabayashi {\it et al}, 
Phys. Rev. B {\bf 59}, 2486 (1999); 
J.~Okabayashi {\it et al}, 
Phys. Rev. B {\bf 64}, 125304 (2001)

\bibitem{krstajic} P.M.~Krstajic {\it et al}, 
Phys. Rev. B {\bf 70}, 195215 (2004)

\bibitem{marel88} D.~van~der~Marel and G.A.~Sawatzky, 
Phys. Rev. B {\bf 37}, 10674 (1988)

\bibitem{solovyev96} I.~Solovyev, N.~Hamada, and
K.~Terakura, Phys. Rev. B {\bf 53}, 7158 (1996)

\bibitem{CaO} I.S.~Elfimov, S.~Yunoki and G.A.~Sawatzky,
Phys. Rev. Lett. {\bf 89}, 216403 (2002)

\bibitem{slater} J.C.~Slater {\it Quantum Theory Of Atomic
Structure} (New York, McGraw-Hill 1960)

\bibitem{Tjeng88} J. Ghijsen {\it et al}, 
Phys. Rev. B {\bf 38}, 11322 (1988)

\bibitem{Altieri_thesis} S.~Altieri {\it Electronic 
Structure of Oxide Thin Films on Metals}, 
Ph.D. thesis, University of Groningen (1999)

\bibitem{Sawatzky_auger_ln} G.A.~Sawatzky, lecture notes on 
Auger spectroscopy

\bibitem{Fuggle} J.C.~Fuggle, in {\it Electron Spectroscopy: 
Theory, Techniques and Applications}, edited by 
C.R.~Brundle and A.~Baker 
(Academic Press, London, 1981), Vol.4, p.85

\bibitem{Tjeng89} L.H.~Tjeng {\it et al} in
{\it Strong Correlation and Superconductivity},
edited by H.~Fukuyama, S.~Maekawa, and A.P.~Malozemoff,
Springer Series in Solid-State Sciences (Springer-Verlag,
Berlin, 1989), Vol.89, p.85

\bibitem{Hao_thesis} L.H.~Tjeng {\it Electronic Structure of 
Oxygen in and on Copper and Silver}, Ph.D. thesis,
University of Groningen (1990)

\bibitem{CrO2} M.A.~Korotin {\it et al},
Phys. Rev. Lett. {\bf 80}, 4305 (1998);
C.F.~Chang {\it et al}, Phys. Rev. B {\bf 71},
52407 (2005)

\bibitem{harrison} W.A.~Harrison: {\it Electronic Structure
and the Properties of Solids}, Dover Publications, Inc., New
York (1989)

\bibitem{kenmochi} K.~Kenmochi, {\it et al}, Jpn. J. Appl. Phys., 
{\bf 43}, L934 (2004)

\bibitem{voogt01} Voogt {\it et al}, Phys. Rev. B {\bf 63},
125409 (2001)

\bibitem{tblmto} O.K.~Andersen, Phys. Rev. B {\bf 12}, 
3060 (1975)

\bibitem{ldau} V.I.~Anisimov, J.~Zaanen and O.K.~Andersen,
Phys. Rev. B {\bf 44}, 943 (1991)

\bibitem{csiszar} S.I.~Csiszar, A.~Rusydi, T.~Hibma,
G.A~Sawatzky, {\it Epitaxial Growth of SrON thin films}, in
preparation

\bibitem{hufner95} Stefan~Hufner, Photoelectron
Spectroscopy, 2nd edition (1995)

\bibitem{st_ratio} K.M.~Siegbahn {\it et al},  
ESCA Applied to Free Molecules, p. 57 (1969)

\end{references}
\end{document}